\documentclass[%
 aip,
 amsmath,amssymb,
 reprint,%
]{revtex4-1}

\usepackage{graphicx}
\usepackage{dcolumn}
\usepackage{bm}

\usepackage[utf8]{inputenc}
\usepackage[T1]{fontenc}
\usepackage{mathptmx}
\usepackage{etoolbox}

\usepackage{placeins}
\usepackage{float}

\makeatletter
\def\@email#1#2{%
 \endgroup
 \patchcmd{\titleblock@produce}
  {\frontmatter@RRAPformat}
  {\frontmatter@RRAPformat{\produce@RRAP{*#1\href{mailto:#2}{#2}}}\frontmatter@RRAPformat}
  {}{}
}%
\makeatother

\usepackage{relsize}
\usepackage{amsmath,amsfonts,amsthm,bm,braket}
\DeclareMathOperator{\Tr}{Tr}
\DeclareMathOperator{\tr}{tr}

\usepackage{float}

\usepackage[caption=false]{subfig}
\usepackage[justification=justified]{caption}
\usepackage{adjustbox}

\usepackage{color} 
\usepackage{soul} 

\usepackage{bbold}
\usepackage{xparse}
\usepackage{comment}

\RenewDocumentCommand{\P}{o g}{%
\IfNoValueTF{#2}{\operatorname{\mathbb{P}}\IfValueT{#1}{_{#1}}}%
{\,{\operatorname{\mathbb{P}}\IfValueT{#1}{_{#1}}}{\bp{#2}}}}

\begin{document}

\preprint{AIP/123-QED}

\title[Determining Quantum Correlation through Nash Equilibria in Constant-Sum Games]{Determining Quantum Correlation through Nash Equilibria in Constant-Sum Games}
\author{A. Lowe}
\email{adam.j.lowe90@gmail.com}
 \affiliation{Department of Applied Mathematics and Data Science, Aston University}

\date{\today}

\begin{abstract}
Quantum game theory has emerged as a promising candidate to further the understanding of quantum correlations. Motivated by this, it is demonstrated that pure strategy Nash equilibria can be utilised as a mechanism to witness and determine quantum correlation. By combining quantum theory with Bayesian game theory, a constant-sum game is designed in which the players are competing against each other, and crucially gain at the other player's expense. Subsequently, it is found that mixed strategy Nash equilibria are only necessary when considering quantum correlation for the designed game. This reveals that a Bayesian game-theoretic framework yields a sufficient condition in which to detect quantum effects.
\end{abstract}

\maketitle

\section{Introduction}\label{s:introduction}
It is of increasing relevance to design interactive experiments which allow for the detection and quantification of quantum correlations \cite{aspect,Passante:2011vz,Aslam:2023vu,Rarity:1990ul,Pelucchi:2022ub}. It has been well-established that quantum mechanics allows scope for advantage in many scenarios, a term which is coined ``Quantum Advantage'' \cite{Hangleiter:2023te,Madsen:2022wo,quantumadvantage,Daley:2022vu}. Since the development of quantum hardware is progressing at a significant rate, it is important to understand under what conditions quantum advantage can be found, and how to classify and understand it. In particular, understanding the strength of quantum correlation and how it can map to quantum advantage is also a key focus of modern technology development \cite{Azuma:2023tz}. Most work has focused on understanding how entanglement and non-locality can be used to witness quantum advantage, displaying significant improvement in recent years \cite{Brunner:2014vw,Tavakoli:2022uf,Li:2023tc}. However, there has been increasing work investigating how quantum correlations beyond entanglement, including quantum discord can further display quantum advantage \cite{Pirandola:2014aa,IP,Lowe:2024aa,lowe,Morimae:2014uh,Knill:1998us,Guha2021quantumadvantage}. This allows the full potential of quantum technologies to be realised by understanding under what circumstances and set-ups quantum mechanics can offer practical advantage beyond traditional entanglement frameworks.

There has been a recent focus on using game-theoretic techniques \cite{tadelis} as inspiration to develop interactive protocols \cite{Ikeda:2023ab}. This enables formal techniques to be utilised to further develop the understanding of the underlying correlation. Game theory has been used as a mathematical framework \cite{nasheq,gamebook} to describe various economic, social and biological interactions by successfully predicting how interactions emerge and why certain phenomena are seen in everyday life \cite{Farooqui:2016aa,Bayes_auctions}. It is this real-world application which makes it an ideal framework in which to consider how quantum effects can modify everyday life, and crucially, determine where quantum mechanics offers a tangible advantage which can be implemented in future quantum technologies \cite{Pappa:2015aa,Roy:2016aa,Brunner:2013aa,Andronikos:2022aa}. Game theory has many subdisciplines which allow for the mathematical frameworks to be tailored to the situation which is described. For example, sequential interactions can be modelled as extensive form games or repeated games \cite{Ikeda:2023aa}. Whereas modelling uncertainty can be achieved best through a Bayesian game-theoretic framework \cite{harsanyi}.

Given there is a requirement to develop protocols which can utilise quantum hardware to witness quantum advantage, the combination of quantum mechanics and game theory has merged into an interdisciplinary field for the development of quantum technologies, and is subsequently known as quantum game theory \cite{PhysRevLett.82.1052,PhysRevA.65.022306,QgameRev,QgameRev1,QGame_Locality,PHOENIX2020126299}. In particular, quantum game theory offers both inspiration and formal techniques with which to solve real-world problems which will arise in a rapidly developing quantum world. This work studies how game-theoretic techniques yield insight into quantum correlations, and in particular, how they can affect strategies of a player, when they are part of an interacting network \cite{cite-key,10.1063/5.0204288}. This is achieved by linking fundamental game-theoretic concepts such as Nash equilibria to differing types of quantum correlation, henceforth understanding the resulting effects on the Nash equilibria as a consequence of the quantum correlation.

\section{Preliminaries}

\subsection{Quantum Discord}
In order to quantify quantum correlation and how to witness and determine quantum effects, a reliable metric needs to be introduced.
Quantum discord measures quantum correlation which can exist beyond the standard techniques of entanglement and non-locality \cite{discord,discord1}. One specific information-theoretic identifier of quantum discord is the difference in the correlation before and after a measurement on a system. This allows the effect of measurement as a quantum process to be deduced, and thus any quantum effects which would only be revealed through this measurement process. The pre-measurement mutual information is the Kullback Leibler divergence between the joint distribution of variables in a coupled system, and the marginal distributions in each of the component subsystems. This may be expressed in terms of the joint entropy of the whole system, $S(\rho^{AB})$ and the marginal entropies $S(\rho^{A})$, $S(\rho^{B})$ of the subsystems $A$ and $B$ as
\begin{equation}
I(\rho^{AB}) = S (\rho^A) + S(\rho^B) - S(\rho^{AB}),
\end{equation}
where, $\rho^{AB}$ is the whole system density matrix and $\rho^{A}$, $\rho^{B}$ are the density matrices calculated as partial traces over the respective subsystems, explicitly $\rho_i = \tr_j \rho_{ij}$ for $i,j \in \{A,B\}$ and $i \neq j$. In terms of the density matrices, the entropies are explicitly calculated as  $S(\rho^{AB}) = - \tr \rho^{AB} \log \rho^{AB} = - \sum_i \lambda_i \ln \lambda_i$, where  $\lambda_i$ are the eigenvalues of this density matrix.

To correctly account for the effect of measurement, a post measurement state is introduced, where the sum over all local measurements is taken. Without loss of generality, the measurement is performed in the $B$ subsystem.
We consider the archetypal case of locally (ie in one of the subsystems) measuring spin up or spin down.   The (conditional) density matrix for subsystem $A$  after a measurement process in $B$ yields a (conditional) density matrix  $\rho^{A | \Pi_{\sigma | \bf{n}}} = \frac{1}{p_{\sigma}} \tr_{B} (\mathbb{1} \otimes \Pi_{\sigma | \bf{n}}) \rho^{AB}  (\mathbb{1} \otimes \Pi_{\sigma | \bf{n}})$, with $\Pi_{\sigma | {\bf{n}}} = \frac{1}{2}(\mathbb{1}+\sigma {\bf{n}}.\boldsymbol{v})$, $\sigma=\pm 1$ denotes spin up/down, ${\bf{n}}=(n_x,n_y,n_z)$ is the Bloch vector, $\boldsymbol{v}=(\sigma_x,\sigma_y,\sigma_z)$ is the vector of Pauli matrices, and $p_{\sigma} = \tr_{B} (\mathbb{1} \otimes \Pi_{\sigma | \bf{n}}) \rho^{AB}$. The pointwise conditional entropy is just 
$S(\rho^{A | \Pi_{\sigma | \bf{n}}})$.

The conditional entropy of $A$ given the measurement in subsystem $B$ is the expectation over the pointwise conditional entropy,
\begin{equation}
S(\rho^{{A} | \Pi_{\sigma|\bf{n}}}) = \sum_{\sigma} p_{\sigma} S(\rho^{A | \Pi_{\sigma | \bf{n}}} ).
\end{equation}
Therefore, the quantum mutual information is now the difference between the marginal entropy of $A$ before measurement and the conditional entropy of $A$ after measurement is
\begin{equation}
J_B (\rho^{AB}) =   S(\rho^A) -  S(\rho^{{A} | \Pi_{\sigma|\bf{n}}}).
\end{equation}

Quantum discord (for subsystem $A$) is the minimum value over all possible measurements of the difference between the pre-measurement mutual information and the post measurement quantum mutual information is defined as
\begin{equation}
\begin{split}
D_A (\rho^{AB}) &=  \min\limits_{\Pi_{\sigma | \bf{n}}} [ I(\rho^{AB}) - J_B (\rho^{AB})] \\&= \min\limits_{\Pi_{\sigma | \bf{n}}} S(\rho^{A | \Pi_{\sigma | \bf{n}}}) + S(\rho^B) - S(\rho^{AB}).
\end{split}
\end{equation}
Note that the minimisation ensures any quantum effects are due to the underlying state, and not due to a specific measurement. Also for a classical state, the post measurement quantum mutual information reduces to the standard classical mutual information, ensuring that the discord is zero for a classical state. Discord is generally asymmetric between subsystems, and the optimisation over all local measurements makes the calculation of discord computationally expensive as the system size increases. A key aspect of discord is its relative robustness to noise \cite{Werlang:2009aa}, which further motivates its potential benefit for practical quantum technologies, as it is less sensitive to perturbations than entanglement.

\subsection{Game Theory}

Game Theory, being the branch of mathematics that deals with decision making by rational players seeking to optimise their own payoffs, with due regard to the likely decision spaces of their opponents, provides an ideal framework for exploring quantum advantage. In particular, the subset of games for which at least one of the players has incomplete information on either of their opponents  or which game they are playing (the class of so called Bayesian games) is appropriate for quantum systems which are physically correlated; for example interacting through entangled photon pairs \cite{PhysRevResearch.6.023248,doi:10.1126/science.abe8770}. 

In the same way that spacelike-separated observers making measurements on quantum correlated physical systems do so in the absence of full knowledge of the other observer's outcomes of measurements, Bayesian game players have to make optimal decisions with only partial knowledge of their opponent's likely strategic choices. Exploiting awareness of this uncertainty, or limited knowledge of the full underlying system,  in both cases underpins optimal decision making. Both instances provide an epistemological perspective of decision making based on what a player is allowed to know about a system, rather than the actual state of the system itself. 

A specific quantum correlated situation is explored that may be posed as a finite sum Bayesian game, the resolution of which in terms of Nash equilibria, will reveal how the game-theoretic solutions can expose the quantum correlations in a non-entangled, but discorded state.

\section{Constant-Sum Bayesian Game}\label{s:Constant-Sum Bayesian Game}

A generic strategic form game  $G$ can be defined as $ G = \{\bm{P}, \Sigma , \bm{U} \}$
with $\bm{P}= \{ P_{1} , P_{2}, \dots, P_{N} \}$ being the set of $N$ players of the game,  The strategy space of the players is $\Sigma = \Sigma _{1} \times \Sigma_{2} \times \dots \times \Sigma_{N}$, where $\Sigma_{i}$ is player $i$'s strategy subspace, and the strategies within player $i$'s respective strategy space are denoted by ${\bm{s}}^{i} \in \{ s_{1}^{i}, \dots, s_{M_{i}}^{i} \}$, where $M_i$ is the number of strategies for player $i$. Then each player has a set of payoffs, $\hat{U}_{i} ( {\bm{s}}^{i}, {\bm{s}}^{-{\bm{i}}})$ where ${\bm{s}}^{-{\bm{i}}}$ denotes the strategies of all the other players, and the set of payoffs themselves depend on all of the players' choices of strategy. 

In games of incomplete information, the uncertainty is incorporated into the game by introducing a distribution over a set of ``types'' of each player. That is to say, each player forms a prior belief over what type of opponent they are facing, which is equivalent to placing a prior belief over a set of different games they might be playing, with different payoffs depending on their and their opponent's choices. 

Formally, a set of types is introduced for each player $i$: $t _i \in T_{i}$, where the prior beliefs are incorporated as a joint probability distribution over types $\P(t_1,  \dots , t_{N})$. The correlations between the players are introduced by a conditional probability distribution $\P( s^{1}_{1}, \dots ,  s^{1}_{M_{1}}, \dots, s^{N}_{1}, \dots , s^{N}_{M_{N}} | t_1,\dots, t_{N})$ conditioned on the unknown types and strategies of all players.

The specific quantum games that will be explored can be phrased as \textit{constant-sum Bayesian games}  in which there are two adversarial players.  The game is designed such that one player ``wins'' only at the expense of the other;  namely $\hat{U}_{1} ( {\bm{s}}^{1}, {\bm{s}}^{2}) = -\hat{U}_{2} ( {\bm{s}}^{1}, {\bm{s}}^{2})$, where $\hat{U}_{1}$ denotes player 1's payoff matrix, and ${\bm{s}}^{1}$ reflects player 1's set of strategies. Similarly for player 2. Note that the sum of the payoffs is a constant (which is often rescaled to be zero, as for the example above). 
Qualitatively, the game can be viewed as a modification of the (cooperative) CHSH game \cite{chsh,Cleve:2004tw} but changed into an adversarial CHSH game.  In this revised game  there are two players, and each can play one of two types of game, where each player has their respective sets of payoffs. Physically, this can be implemented by a referee who sends each player (now termed Alice and Bob for convenience), one part of a shared state. At this stage, no assumption is made on the type of correlation of the shared state. Then each player, can perform a measurement on their part of the shared state, and obtain an outcome, either $\pm 1$ (which conveniently maps over to spins of a particle, as before for the definition of discord). Each player then informs the referee about their measurement, who then assigns each player a payoff based on their measurement and on which type of game was played. The Bayesian nature of the game is due to the lack of knowledge each player has about the other's type, and hence payoff matrix, requiring a prior distribution over types to be introduced. 

In a physical experiment, this implies each player has no knowledge of what detector the other player will use to measure the observable, and using a given detector alters the assigned payoffs. This is explicit in Table \ref{tab1}, which uses the notation that $\alpha \in \{ {\bm{a}},{\bm{a'}}\}$ corresponds to Alice's types of detectors, where $\bm{a}$ is one detector, and $\bm{a'}$ is the other detector. Similarly for Bob $\beta \in \{ {\bm{b}},{\bm{b'}}\}$. The measurement outcomes have been assigned figuratively as $\uparrow, \downarrow$, depending on what `spin' is measured in the physical system. The set of payoffs are presented as a matrix for each player.
\begin{table}
\Large
    \centering
    \begin{tabular}{c|c c}
       ($\alpha$,$\beta$)  & ${\bm{b}}$ &${\bm{b'}}$ \\ \hline
     ${\bm{a}}$  & \begin{tabular}{c|c c}
       &  {$\uparrow$} &{$\downarrow$}\\ \hline
       {$\uparrow$}  & 1,0 &0,1  \\
      {$\downarrow$} & 0,1 & 1,0
     \end{tabular}  & \begin{tabular}{c|c c}
       & {$\uparrow$} & {$\downarrow$} \\ \hline
       {$\uparrow$} & 1,0 &0,1  \\
       {$\downarrow$} & 0,1 & 1,0
     \end{tabular} \\ 
     ${\bm{a'}}$  & \begin{tabular}{c|c c}
       &  {$\uparrow$} & {$\downarrow$} \\ \hline
        {$\uparrow$} &  1,0 &0,1  \\
        {$\downarrow$} & 0,1 & 1,0
      \end{tabular} & \begin{tabular}{c|c c}
       &  {$\uparrow$} & {$\downarrow$} \\ \hline
       {$\uparrow$} & 0,1  & 1,0 \\
        {$\downarrow$} & 1,0 & 0,1
      \end{tabular}
    \end{tabular}
    \caption{The first number denotes Alice's payoff ($U_A$), and the second number denotes Bob's payoff ($U_B$). Based on what bits they receive, ${\bm{a}},{\bm{b}},{\bm{a'}},{\bm{b'}}$ denote how Alice and Bob's detectors are set up. From this, they then perform measurements on their shared states and based on their results, they are assigned a payoff.}
        \label{tab1}
\end{table}
 
\subsection{Expected Payoff} 
To compute the overall expected payoff for each player in this scenario, the averaging must account for each player's respective prior beliefs, the conditional probability distribution between the players, and each player's local tensor of payoffs. Explicitly, the overall expected payoff for player $i$ is
\begin{equation}
\label{e:cfn}
U_i = \sum_{\alpha, \beta} \sum_{\sigma, \sigma'} U^{\alpha,\beta}_{\sigma,\sigma',i} \P(\sigma,\sigma'|\alpha ,\beta)\P_i(\alpha,\beta),
\end{equation}
where $i$ denotes the given player,
$\alpha \in \{ \bm{a},\bm{a'} \},
\beta \in  \{ \bm{b},\bm{b'} \},
$
$U^{\alpha,\beta}_{\sigma,\sigma',i}$ is the tensor of payoffs, $\P_i(\alpha,\beta)$ are player $i$'s prior beliefs, and $\sigma,\sigma' \in \{1,-1\}$ where $1$ represents measuring spin up, and $-1$ represents spin down.
Additionally, the conditional probability distribution is given by
\begin{equation}
\label{condProb}
 \P(\sigma,\sigma'|\alpha ,\beta) = \Tr[\Pi_{\sigma|\alpha} \otimes \Pi_{\sigma'|\beta} \rho],
\end{equation}
where
\begin{equation}
\Pi_{\sigma| {\alpha}}
\equiv \frac{1}{2} [1+\sigma \alpha.\boldsymbol{v}],
\end{equation}
and $\rho$ is the shared state between the players.

The elements of $\alpha$ are parameterised on the Bloch sphere, as before for the computation of quantum discord. This ensures the players' strategies are choices of angles on the Bloch sphere. For simplicity, it is assumed that the spherical angles are set to zero, leaving only the polar angles to be chosen.
This conditional distribution accounts for the correlation between the players, and subsequently can be used to determine whether quantum correlation can be used as a mechanism for quantum advantage.

 The referee determines whether $\bm{a}$ or $\bm{a'}$  is to be used. Respectively for $\bm{\beta}$, where each corresponds to which one of a pair of measuring devices is used. 
 Note, $\bm{a} \equiv \bm{a}(\theta_{a})$, $\bm{a'} \equiv \bm{a'}(\theta_{a'})$ and similarly for the elements in $\beta$. Crucially, the players get to choose the measurement angles in the apparatus $\{\theta_{a},\theta_{a'} \}$.  So for the pure strategies of choosing one or the other of each of the two choices, a measurement is made and the result is forwarded to the referee who determines the expected payoff returned to each player based on their strategies and the decomposition of the joint distribution into their respective projection spaces.
Expanding (\ref{e:cfn}) explicitly, and suppressing the player index, the result is
\begin{align}
\label{e:cfn2}
\begin{split}
U &=
\sum_{\sigma, \sigma'} U^{\bm{a},\bm{b}}_{\sigma,\sigma'} \P(\sigma,\sigma'|\bm{a} ,\bm{b})\P(\bm{a},\bm{b}) \\&+ \sum_{\sigma, \sigma'} U^{\bm{a},\bm{b'}}_{\sigma,\sigma'} \P(\sigma,\sigma'|\bm{a} ,\bm{b'})\P(\bm{a},\bm{b'})\\
{}&+ \sum_{\sigma, \sigma'} U^{\bm{a'},\bm{b}}_{\sigma,\sigma'} \P(\sigma,\sigma'|\bm{a'} ,\bm{b})\P(\bm{a'},\bm{b}) \\&+ \sum_{\sigma, \sigma'} U^{\bm{a'},\bm{b'}}_{\sigma,\sigma'} \P(\sigma,\sigma'|\bm{a'} ,\bm{b'})\P(\bm{a'},\bm{b'}).
\end{split}
\end{align}
At this point, it is assumed that the priors are all the same and can be absorbed into definitions of `local' utility $U^{\alpha,\beta}_{\sigma,\sigma'}$. The physical justification for this is that each player has no knowledge of which type of detector they, or the other player will be using. Therefore, it is assumed that $\P({\bm{a},\bm{b}}) = \P({\bm{a},\bm{b'}})  = \P({\bm{a'},\bm{b}})  = \P({\bm{a'},\bm{b'}})  = 1/4$, since $\P({\bm{a},\bm{b}})$ is an independent joint probability, where $\P({\bm{a}}) = \P({\bm{b}}) = 1/2$. Similarly for ${\bm{a'}}$ and ${\bm{b'}}$.

\subsection{Determining Pure Strategy Nash Equilibria}
To develop insight between Nash equilibria and quantum correlation, the Nash equilibria are explored for the considered game. Since it is a Bayesian game, where each player has two strategic choices, there are four simultaneous Nash equilibrium equations which must be satisfied to ensure each player has no incentive to unilaterally deviate,
\begin{align}
U_A (\theta_a, \theta_{a'},\theta_b,\theta_{b'}) &\geq U_A(\theta_{a}^{*}, \theta_{a'},\theta_b,\theta_{b'}), \label{nash1} \\
U_A (\theta_a, \theta_{a'},\theta_b,\theta_{b'}) &\geq U_A(\theta_{a}, \theta_{a'}^{*},\theta_b,\theta_{b'}), \\
U_B (\theta_a, \theta_{a'},\theta_b,\theta_{b'}) &\geq U_B(\theta_{a}, \theta_{a'},\theta_{b}^{*},\theta_{b'}), \\
U_B (\theta_a, \theta_{a'},\theta_b,\theta_{b'}) &\geq U_B(\theta_{a}, \theta_{a'},\theta_b,\theta_{b'}^{*}),
\end{align}
where $\theta_{\alpha(\beta)}^{*}$ is an arbitrary alternative strategy choice and $U_{A(B)}$ denotes Alice's (Bob's) expected payoff. Therefore, each player wishes to choose a set of strategies where there is no incentive to deviate, as when in Nash equilibrium, any deviation will result in a worse or no better  expected payoff.
It should be emphasised that for the scenario considered, the players are choosing their measurement angles with a given pure strategy. Explicitly, this means the players are only choosing one angle in a given type of game, not a statistical mixture of angles. This of course reflects the physical situation of having to choose an actual angle for the measurement device in each experiment. It is unphysical to consider a mixture of choices of angles in a single-run of an experiment. However, it is an interesting hypothetical issue as to how the players choosing a statistical mixture of angles over repeated iterations of the experiment affects the Nash equilibria in a combative game, but this is beyond the scope of this work. 

Explicitly analysing Eq. (\ref{nash1}), gives further insight on equilibria. Consider, for example, the equilibrium payoff,
\begin{equation}
\begin{split}
 &\sum_{ \bm{a}, \beta} \sum_{\sigma, \sigma'} U^{ \bm{a},\beta}_{\sigma,\sigma'} \P(\sigma,\sigma'| \bm{a} ,\beta)\P( \bm{a},\beta) \\&\geq  \sum_{ \bm{a}^{*}, \beta} \sum_{\sigma, \sigma'} U^{ \bm{a}^{*},\beta}_{\sigma,\sigma'} \P(\sigma,\sigma'| \bm{a}^{*} ,\beta)\P( \bm{a}^{*},\beta).
 \end{split}
\end{equation}
Recall  that $\bm{a} \equiv \bm{a}(\theta_{a}) \implies \bm{a}^{*} \equiv \bm{a}(\theta_{a}^{*})$ and noting that the tensor of payoffs is independent of the measurement, therefore $ U^{\bm{a},\beta}_{\sigma,\sigma'}= U^{\bm{a}^{*},\beta}_{\sigma,\sigma'}$. 
 Additionally, it is assumed that the players are not aware of what type of game they are in, so this also does not affect the measurement, subsequently $\P(\bm{a},\beta) = \P(\bm{a}^{*},\beta)$.
This allows the equilibrium payoffs to be re-arranged to yield
\begin{equation}
\begin{split}
\label{eq:ineq1}
 &\sum_{\bm{a}, \beta} \sum_{\sigma, \sigma'} U^{\bm{a},\beta}_{\sigma,\sigma'} \P(\bm{a},\beta) \big[\P(\sigma,\sigma'|\bm{a} ,\beta) \\&- \P(\sigma,\sigma'|\bm{a}^{*} ,\beta) \big] \geq 0.
\end{split}
\end{equation}
Furthermore, since the tensor of payoffs are positive semi-definite and the prior beliefs satisfy the standard axioms of probability theory, without loss of generality, Eq. (\ref{eq:ineq1}) can be written as
\begin{equation}
\P(\sigma,\sigma'|\bm{a},\beta) - \P(\sigma,\sigma'|\bm{a}^{*} ,\beta) \geq 0.
\end{equation}
This is now solely expressed in terms of the players' measurements and correlation,  explicitly,
\begin{equation}
\begin{split}
\label{neweq1}
\Tr[\Pi_{\sigma|\bm{a}} \otimes \Pi_{\sigma'|\beta} \rho] - \Tr[\Pi_{\sigma|\bm{a}^{*}} \otimes \Pi_{\sigma'|\beta} \rho] &\geq 0, \\
=\Tr[\Pi_{\sigma|\bm{a}} \otimes \Pi_{\sigma'|\beta} \rho -  \Pi_{\sigma|\bm{a}^{*}} \otimes \Pi_{\sigma'|\beta} \rho] &\geq 0, \\
=\Tr[(\Pi_{\sigma|\bm{a}} - \Pi_{\sigma|\bm{a}^{*}}) \otimes \Pi_{\sigma'|\beta} \rho] &\geq 0.
\end{split}
\end{equation}

It is clear from Eq. (\ref{neweq1}), that for each of the Nash equilibria equations to be satisfied, the measurements in a given subsystem will control the sign of the inequality.
To fully understand how this relates to the correlation between the players, it is insightful to consider how general classes of states affect the measurement. For non-separable states which violate Bell inequalities, the state will mix the measurement subspaces,  explicitly highlighting the effects of the superposition of the state. For mixed separable states, it is easier to understand how the state affects the measurement subsystem. For example, considering a general mixed state given by
\begin{equation}
\rho = \sum_j q_j \rho_{j}^{A} \otimes \rho_{j}^{B},
\end{equation}
where $0\leq q_j \leq 1$, $\sum_j q_j = 1$, and $\rho_{j}^{A(B)}$ represents Alice's (Bob's) local state in their respective subspace.
Substituting this into Eq. (\ref{neweq1}), gives
\begin{equation}
\begin{split}
&\Tr[(\Pi_{\sigma|\bm{a}} - \Pi_{\sigma|\bm{a}^{*}}) \otimes \Pi_{\sigma'|\beta} \rho] \\&= \sum_j q_j \Tr[(\Pi_{\sigma|\bm{a}} - \Pi_{\sigma|\bm{a}^{*}}) \rho_{j}^{A} \otimes \Pi_{\sigma'|\beta} \rho_{j}^{B}]. 
\end{split}
\end{equation}
Therefore, it is clear that any quantum effects will be due to the local subsystem which the measurements act on, and the fact that there is a statistical mixture of these subsystems. To be clear, considering an entirely classical separable system given by $\rho= \rho_A \otimes \rho_B$, the above result is
\begin{equation}
\begin{split}
&\Tr[(\Pi_{\sigma|\bm{a}} - \Pi_{\sigma|\bm{a}^{*}}) \otimes \Pi_{\sigma'|\beta} \rho] \\&= \Tr[(\Pi_{\sigma|\bm{a}} - \Pi_{\sigma|\bm{a}^{*}}) \rho^{A} \otimes \Pi_{\sigma'|\beta} \rho^{B}],
\end{split}
\end{equation}
which clearly does not contain any statistical mixture of the state.
In order to understand the physical interpretation of how this affects Nash equilibria, it is insightful to consider some exemplar states. Calculating Nash equilibria is an optimisation problem which consists of simultaneously solving four inequalities. Given it is a constant-sum game, this game can be solved using minimax procedures, thus necessitating the need to find saddle point solutions. More precisely, the expected payoffs for the constant-sum game designed for Alice and Bob can be written as
\begin{equation}
\begin{split}
U_A (\theta_a, \theta_{a'},\theta_b,\theta_{b'}) &= C + f(\theta_a, \theta_{a'},\theta_b,\theta_{b'}) \equiv C + f, \\
U_B (\theta_a, \theta_{a'},\theta_b,\theta_{b'}) &= C - f(\theta_a, \theta_{a'},\theta_b,\theta_{b'}) \equiv C - f,
\end{split}
\end{equation}
respectively, where $C$ is the constant of the game. Note for zero-sum games, $C=0$. Therefore, for Alice to maximise their payoff, they wish to maximise the function $f$. Whereas, Bob wishes to minimise $f$ to maximise their payoff, hence the need for minimax solutions. This naturally extends itself to a saddle-point scenario.
The procedure to find the saddle-point of $f$, and thus the minimax solutions is introduced below.

It is important to understand the nature of the function which is to be optimised. Since $f$ is continuous, standard calculus techniques can be utilised in order to find the optimal solution. Specifically, using the definition of the Jacobian as
\begin{equation}
\label{J}
J = \begin{pmatrix} \frac{\partial f}{\partial \theta_a} &  \frac{\partial f}{\partial \theta_{a'}} &  \frac{\partial f}{\partial \theta_b} &  \frac{\partial f}{\partial \theta_{b'}} \end{pmatrix},
\end{equation}
allows the local curvature of $f$ to be determined. To further classify the curvature, and determine whether the nature of the extrema are saddle points, further analysis is required. 
Specifically, the local curvature around these critical points where  $\nabla  f=0$ is determined from the Hessian matrix, $ \hat{H} = \nabla^2 f$ is evaluated at the vector of stationary point solutions, given by ${\bf{y}} = \begin{pmatrix} \tilde{\theta}_{a} & \tilde{\theta}_{a'} & \tilde{\theta}_{b}& \tilde{\theta}_{b'} \end{pmatrix}^{T}$.

\begin{figure*}[htb!]
  \centering
  \subfloat[a][] {\includegraphics[width=0.42\linewidth]{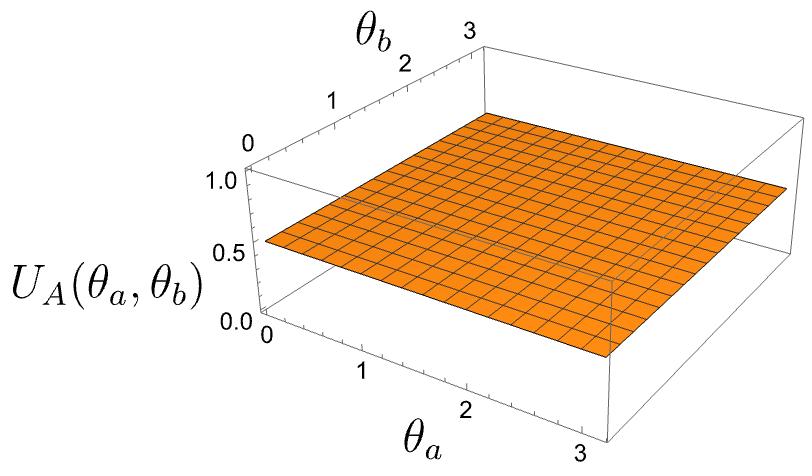}} 
  \subfloat[a][] {\includegraphics[width=0.42\linewidth]{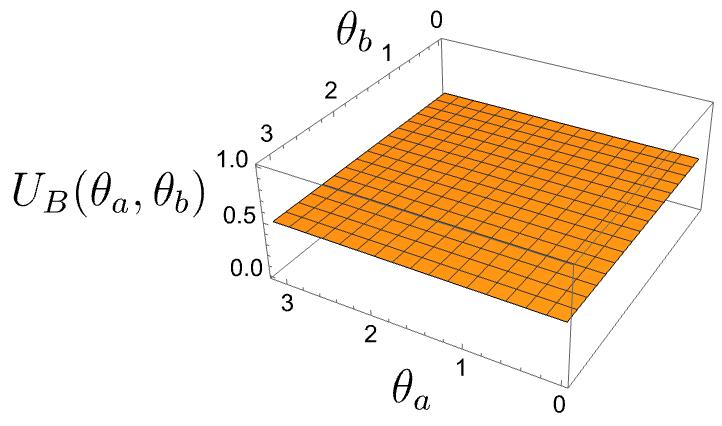}} 
  \\\subfloat[a][] {\includegraphics[width=0.42\linewidth]{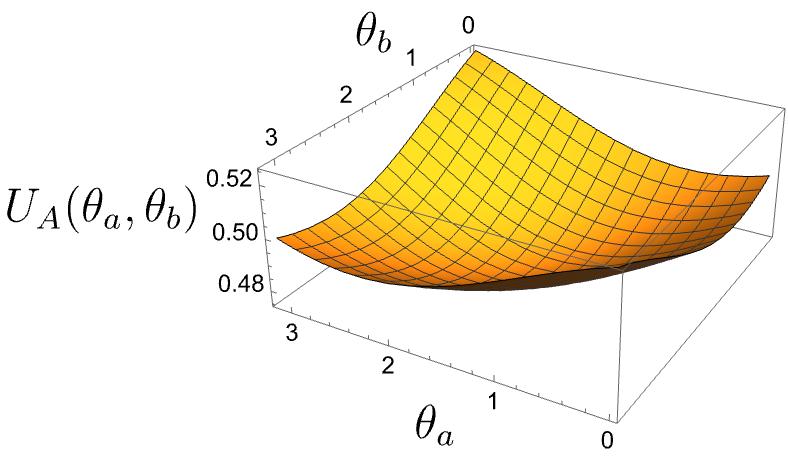}} 
 \subfloat[a][]{\includegraphics[width=0.42\linewidth]{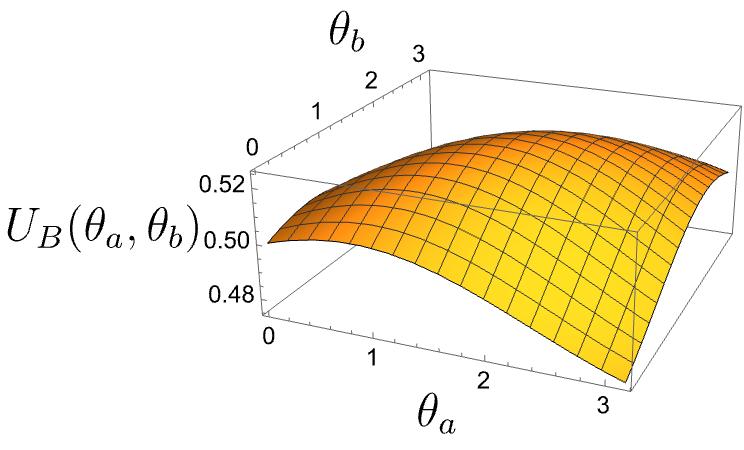}} \\
  \caption{Werner state example: (a), (b) are for classical correlation ($\eta=0$) and (c),(d) for $\eta=0.1$. In (a) and (b), the surface is flat, thus independent of the players' strategies so there is no incentive to deviate resulting in a weak Nash equilibrium. For quantum correlation in subfigures (c),(d), there is no saddle point and hence no Nash equilibria as there are only global minima or maxima. Note, $\theta_{a'} = \theta_{b'} = \pi/2$ is chosen for illustrative purposes.} \label{fig:1}
\end{figure*}
The procedure thus follows the following {\it{Pseudo-Algorithm}}, which details the steps taken to elicit the behaviour of the payoff function.

{\it{Pseudo-Algorithm:}}
\begin{itemize}
\item[1.] Use the initial state to compute the expected payoff for each of the players, and thus determine $f$.
\item[2.] Find the Jacobian of the expected payoff using Eq. (\ref{J}).
\item[3.] Find the stationary points of the function by calculating the roots on the interval $[0,2\pi]$ using the Jacobian.
\item[4.] Compute the Hessian matrix, and find the signs of the diagonal entries to determine nature of the critical points.
\item[5.] Use the signs of the diagonal entries of the Hessian to determine saddle points, and thus Nash equilibria or otherwise.
\end{itemize}
The {\it{Pseudo-Algorithm}} was implemented in Python using numerical optimisation through the Python libraries numpy and scipy. A maximum sampling grid size was chosen where there were approximately 2.8 million points which were sampled over a range of $[0,2\pi]$. For each of these sampled points, it was checked whether there was a saddle point solution. These values were then checked visually using Mathematica, as seen in the Figures throughout. This was confirmed by substituting the found values into the expected payoff for both players.

It should be emphasised that the nature of the saddle points can be determined entirely from the diagonal elements of the Hessian, as they represent the local change in each players specific strategy, namely for Alice, ($\theta_a, \theta_{a'}$) and for Bob, ($\theta_b, \theta_{b'}$) by determining whether the critical points are local maxima or minima. Subsequently, to ensure neither player has any incentive to deviate from their strategy when considering the function $f$, the solution must be a minimum for Bob, and a maximum for Alice. Therefore the leading two diagonal entries must contain a negative semi-definite sign, and the bottom two diagonal entries must contain a positive semi-definite sign. At this stage it is important to clarify, that zero diagonal entries would also satisfy the conditions for Nash equilibria, however these would be classified as weak rather than strict equilibria.

\subsection{Specific States}
To illustrate the relationship between Nash equilibria and discorded states, we now consider explicit, known examples of quantum states which have varying quantum correlation as functions of their respective parameters. For each state introduced, the {\it{Pseudo-Algorithm}} was implemented to determine whether there were saddle points or not by exploring what values for the players' measurements were critical values, and then understanding the curvature around these critical points. 

\subsubsection{Werner State}
\begin{figure*}[htb!]
  \centering
  \subfloat[a][] {\includegraphics[width=0.42\linewidth]{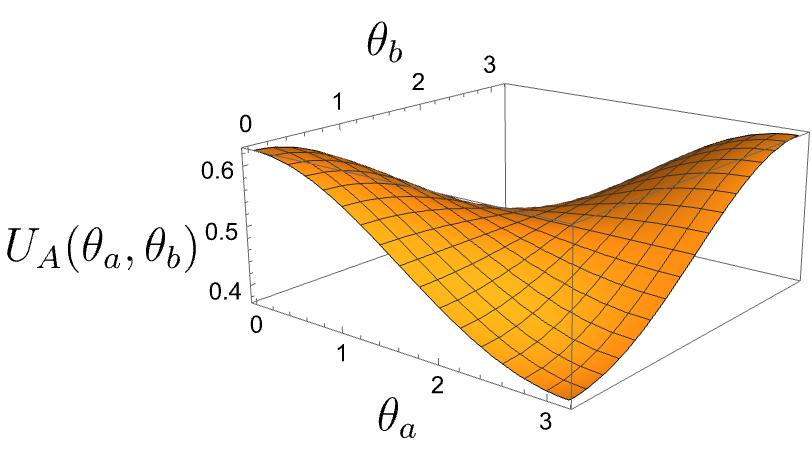}} 
   \subfloat[a][] {\includegraphics[width=0.42\linewidth]{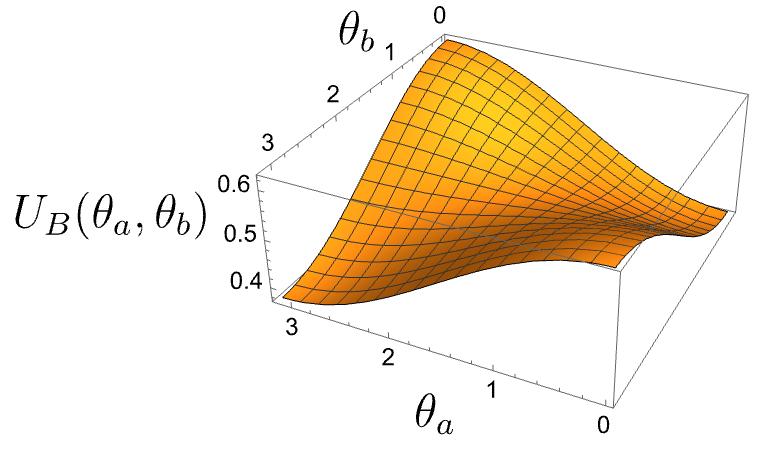}} 
 \\\subfloat[a][] {\includegraphics[width=0.42\linewidth]{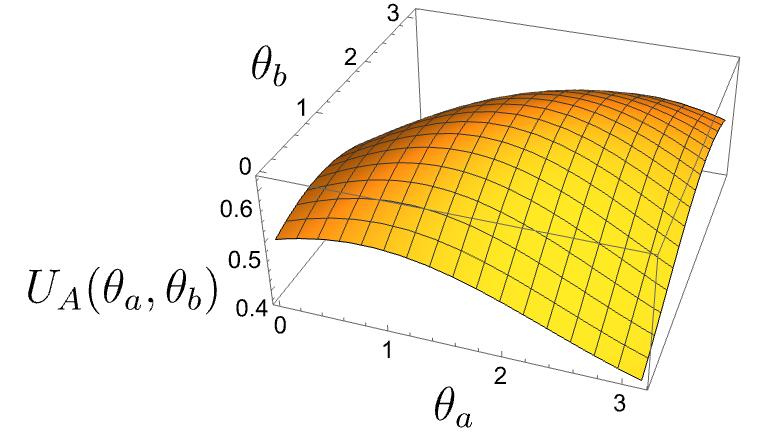}} 
 \subfloat[a][]{\includegraphics[width=0.42\linewidth]{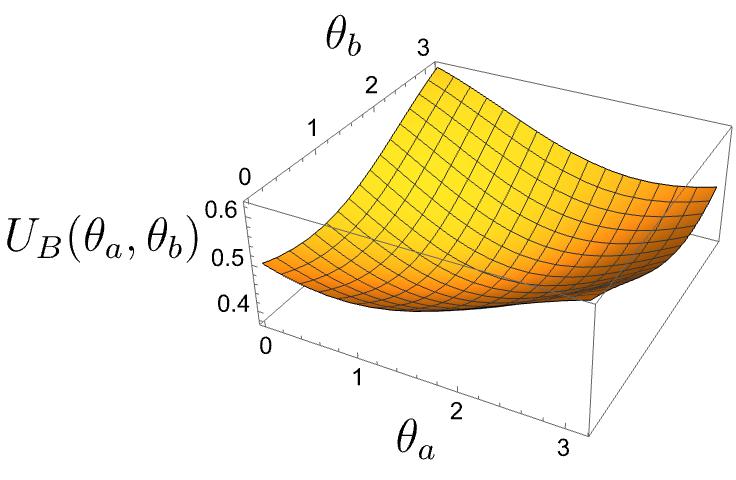}} \\
  \caption{Discord Example: Subfigures (a),(b) are classical correlated ($x=0$).  Subfigures (c),(d) are quantum correlated ($x=\pi/2$).  This demonstrates a clearer picture of a pure strategy solution in (a) and (b) as there is an obvious saddle point giving a strict Nash equilibrium. The saddle point only occurs in the classical regime, with $\theta_{a'} = \theta_{b'} = \pi / 2$. For quantum correlation, there are no longer any saddle point solutions. Subfigures (c) and (d) show the expected payoffs using the optimal parameters found for subfigures (a), (b) showing simple maximum/minimum surfaces.} \label{fig:2}
\end{figure*}
The Werner state is a useful example as it varies from being uncorrelated, quantum discorded, entangled, and finally exhibits Bell non-locality. Thus it is an insightful  example when exploring how specific quantum correlations can impact system behaviour. The Werner state is given by
\begin{equation}
\rho_W (\eta) = \frac{\mathbb{1}- \eta}{4} + \eta \ket{\psi_{-}} \bra{\psi_{-}},
\end{equation}
where $\ket{\psi_{-}} = \frac{1}{\sqrt{2}}(\ket{\uparrow \downarrow} - \ket{\downarrow \uparrow})$, and $\eta \in [0,1]$. This state is known to be classically correlated for $\eta=0$, discorded but non-entangled in the region $0 < \eta \leq 1/3$, and entangled otherwise. 
Considering the classical scenario when $\eta=0$, this implies that the state becomes maximally mixed, explicitly $\rho_{W}(0) = \frac{\mathbb{1}}{4}$. Substituting the Werner state into Eq. (\ref{condProb}) gives the conditional probability as
\begin{equation}
\begin{split}
 \P(\sigma,\sigma'|\alpha ,\beta) &= \Tr[\Pi_{\sigma|\alpha} \otimes \Pi_{\sigma'|\beta} \rho_{W}(0)] \\&= \frac{1}{4}  \Tr[\Pi_{\sigma|\alpha} \otimes \Pi_{\sigma'|\beta}] 
\\&= \frac{1}{4}  \Tr[\Pi_{\sigma|\alpha}] \otimes \Tr[\Pi_{\sigma'|\beta}]= \frac{1}{4},
\end{split}
\end{equation}
since the trace of two products is the product of two traces, and the trace of the projective measurements is $1$. Note that this no longer depends on the players' measurements which will result in the expected payoff becoming a constant whenever the Werner state is classical, as expected. This is seen explicitly in Fig. \ref{fig:1}. However, as the parameter $\eta$ is increased, the previous analysis is no longer valid, thus the players' measurements can affect the expected payoff. It is found that once this occurs for the Werner state, there is then no saddle point for this function, thus ensuring there are no pure strategy Nash equilibria. Since when $\eta = 0$, the expected payoff is independent of the measurement, any change to the measurement does not change the expected payoff, subsequently there is no incentive to deviate. Therefore, this is a weak Nash equilibrium. Furthermore, it was checked for increasing values of $\eta$ using the {\it{Pseudo-Algorithm}} that no saddle point emerged. In particular it was checked for the non-local region when $\eta=1$, which is when the Werner state reduces to a Bell state.


\subsubsection{Zero entanglement, non-zero discord state}

The question arises as to how mixed separable states affect the Nash equilibrium. It is clear that if the Nash equilibria are changed by a mixed separable state, then quantum effects have a profound effect on implementing optimal strategic choices in a game-theoretic setting. 
To further understand how quantum discord differs from classical correlation, a discorded state which exhibits no entanglement is introduced,
\begin{equation}
\rho_{D_1} (x) = \frac{1}{2} \Big[ \ket{\uparrow} \bra{\uparrow}  \otimes \ket{\uparrow} \bra{\uparrow} + \ket{x} \bra{x} \otimes \ket{x} \bra{x}\Big],
\label{D1}
\end{equation}
where $\ket{x} = \cos \frac{x}{2} \ket{\uparrow} + \sin \frac{x}{2} \ket{\downarrow}$ and $x \in [0,2\pi]$. This is clearly a mixed separable state, which has classical correlation at $x=0,\pi$, and has quantum discord otherwise. For this state, the measurement always impacts the expected payoff as can be seen in Fig. \ref{fig:2}. Once again, it is seen that when the state is classical, there is a saddle point, and thus a pure strategy solution. No pure strategy solution occurs in the quantum regime, as both players will always have incentive to deviate irrespective of what the other player chooses to measure. This shows a consistent qualitative understanding to the results for the Werner state. Again, this discorded state was checked for different values of $x$ which yield quantum correlation, demonstrating the robustness of this analysis.

\subsubsection{Quantum-Classical state}
\begin{figure*}[htb!]
  \centering
  \subfloat[a][] {\includegraphics[width=0.42\linewidth]{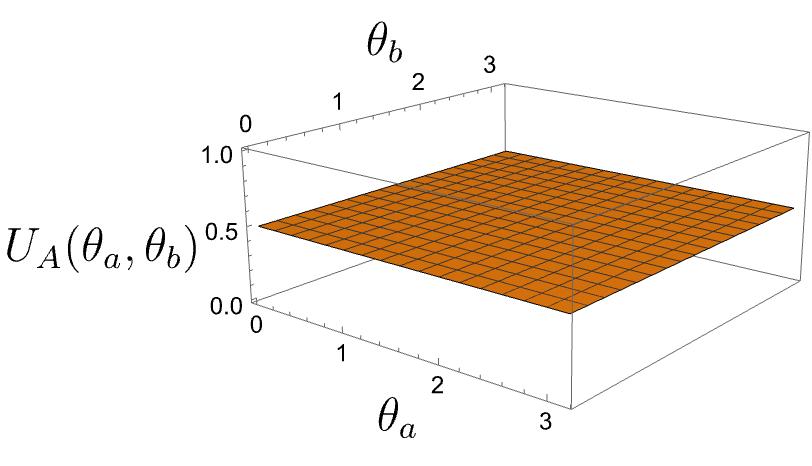}} 
   \subfloat[a][] {\includegraphics[width=0.42\linewidth]{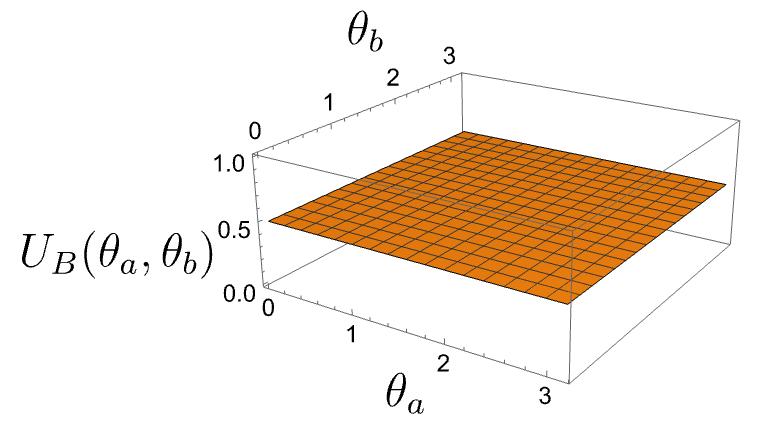}} 
 \\\subfloat[a][] {\includegraphics[width=0.42\linewidth]{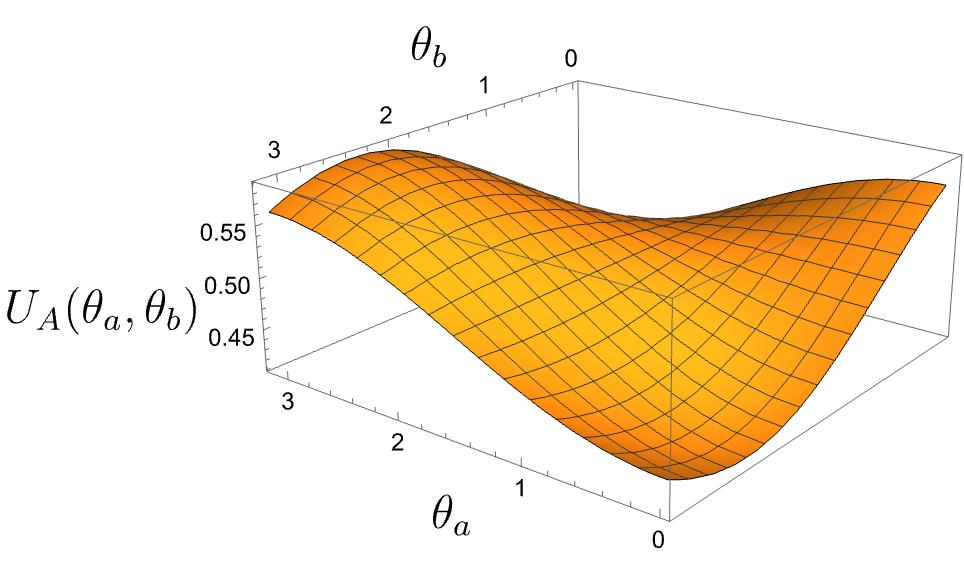}} 
 \subfloat[a][]{\includegraphics[width=0.42\linewidth]{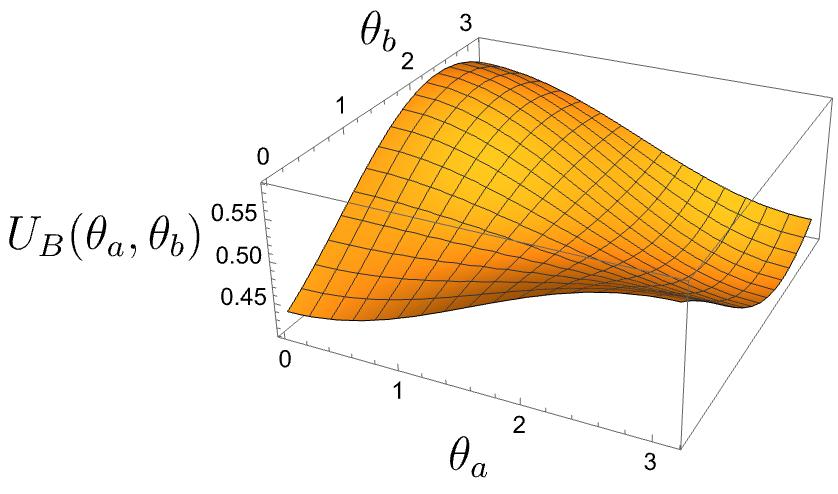}} \\
  \caption{One-way quantum-classical discord. (a),(b) with $x=0$ (classical). (c),(d) with $x=\pi/2$ (quantum).  Parts (a) and (b) are weak Nash equilibria due to the flat surface, whereas (c) and (d) are found to have fixed strategies which yield saddle points for $\theta_{a'}=\pi/2$ and $\theta_{b'} = \pi/4$. Nash equilibria persist for quantum and classical states.} \label{fig:3}
\end{figure*}
The states considered so far are both symmetric, namely the discord is the same irrespective over which subsystem is optimised over. However, there are discorded states which exhibit quantum effects when one subsystem is measured, but exhibits classical effects when the other subsystem is measured. Therefore, it is not clear how the Nash equilibria will be affected by these types of discorded states. Subsequently, a final example is an asymmetric state which exhibits one-way quantum discord (i.e. one player possesses quantum correlation, the other player only has access to classical correlation), given by
\begin{equation}
\rho_{D_2} (x) = \frac{1}{2} \Big[ \ket{\uparrow} \bra{\uparrow}  \otimes \ket{\uparrow} \bra{\uparrow} + \ket{\downarrow} \bra{\downarrow} \otimes \ket{x} \bra{x}\Big].
\label{D2}
\end{equation}
This state exhibits similar qualitative behaviour to the Werner state when both states are classical.  However the reason for this similar behaviour is due to different causes.  For the Werner state, the existence of weak Nash equilibria was due to the state being maximally mixed.  For this discorded state the payoff behaviour is due to a cancellation when calculating the payoffs between the players. Explicitly, for the two classical cases ($x=0$, $x=\pi$) 
\begin{equation}
\begin{split}
 \P(\sigma,\sigma'|\alpha ,\beta) &= \Tr[\Pi_{\sigma|\alpha} \otimes \Pi_{\sigma'|\beta} \rho_{D_2}(0)] \\&= \frac{1}{4} (1+\sigma' \cos \theta_{\beta}),
\end{split}
\label{cancel1}
\end{equation}
and 
\begin{equation}
\begin{split}
 \P(\sigma,\sigma'|\alpha ,\beta) &= \Tr[\Pi_{\sigma|\alpha} \otimes \Pi_{\sigma'|\beta} \rho_{D_2}(\pi)] \\&= \frac{1}{4} (1 + \sigma \sigma' \cos \theta_{\alpha} \cos \theta_{\beta}),
\end{split}
 \label{cancel2}
\end{equation}
 highlighting the asymmetry in correlations due to the state in Eq. (\ref{D2}).  For the states where the correlations depend only on one player's measurement, the cosine terms cancelled out in Eq. (\ref{cancel1}) when the expected payoff was computed. Subsequently, the expected payoff becomes a constant, which explains the weak Nash equilibrium surfaces in subfigures (a),(b) in Fig. \ref{fig:3}. When computing the expected payoff for Eq. (\ref{cancel2}), there is no cancellation in the payoffs so the expected payoff has a surface qualitatively similar to Fig. \ref{fig:2} with strict Nash equilibria saddle points. It is also interesting to note that the final two states can be created through local operations and classical communication (LOCC), demonstrating how the interaction prior to a game can drastically alter the strategic optimisation when solving the game. Explicitly, the LOCC can ensure that a given subsystem is mixed through certain operations on the subsystem, whereas the other subsystem remains classical.
 In this case for the genuine quantum scenario of $x=\pi/2$ there also exists a strict Nash equilibrium saddle point as shown in Fig \ref{fig:3}, parts (c),(d).

It is speculated that the saddle points persisting in the quantum case in Fig. \ref{fig:3} are likely due to the state being classical-quantum, i.e. one of the subsystems is entirely classical. In terms of quantum discord, this means that asymmetry arises as when it is computed with a measurement on one subsystem the discord will be non-zero, but if the measurement is performed on the other subsystem, it will be zero. This further suggests that quantum behaviour is affecting the Nash equilibria, as any form of classical correlation appears to yield pure strategy Nash equilibria.

\subsubsection{Biased game with Quantum-Classical state}
\begin{figure*}[htb!]
  \centering
  \subfloat[a][] {\includegraphics[width=0.4\linewidth]{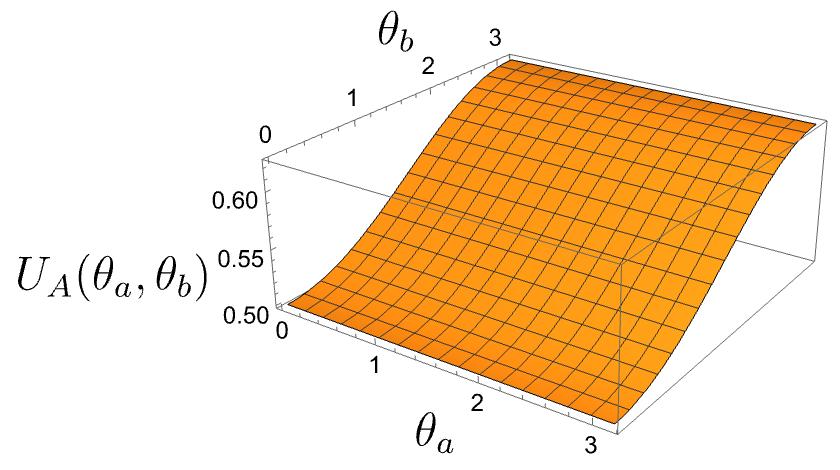}} 
  \subfloat[a][] {\includegraphics[width=0.4\linewidth]{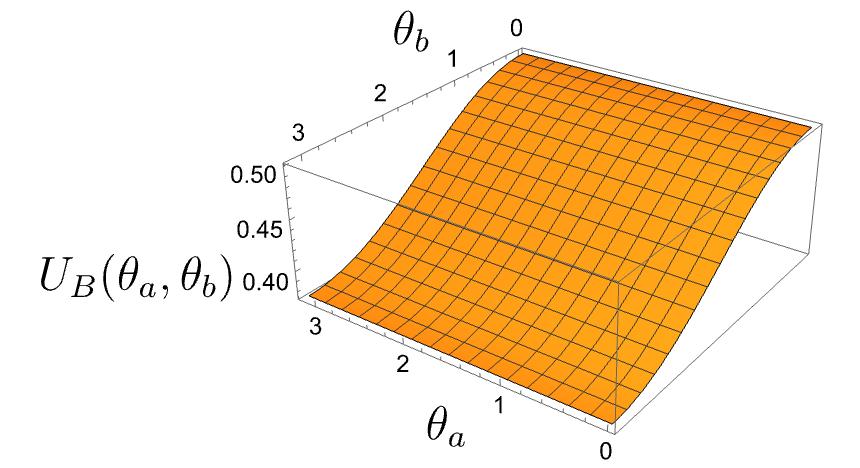}} 
  \caption{Biased (asymmetric) Payoff Game for the quantum-classical state of Figure \ref{fig:3}, again with $x=0$ (classical) for subfigures (a),(b).  Now (a) and (b) have measurement-dependence for one of the players unlike in the unbiased game, ensuring the optimisation is a maximisation of the payoffs for one of the players (the other player's measurements have no influence on the payoffs). Again it is found that there are Nash equilibria for both the classical and quantum scenarios.} \label{fig:4}
\end{figure*}
Given that the expected payoff is constant regardless of the choice of the classical player when considering $\rho_{D_2} (0)$, and noting this is due to the symmetry of the payoffs, it is interesting to introduce asymmetry into the payoffs and understand how this affects the expected payoff. For example, we can create a biased, but still constant-sum game if we modify the payoff matrix in Table \ref{tab1} where the payoff entry at $u_{\downarrow,\downarrow}^{{\bf{a}},{\bf{b}}}$ substitutes   $1$ by $2$ for Alice, and for Bob, the payoff changes from $0$ to $-1$.  Note that this is still a constant-sum game, however now the game is biased towards Alice. For this set of decision payoffs, the bias destroys the symmetric cancellation of measurement effects and has dependence on Bob's measurement for $x=0$. Thus the optimisation now becomes a maximisation decision for Bob, though Alice has no control over the payoff. This is seen explicitly in Fig. \ref{fig:4}.


\section{Discussion}\label{s:Discussion}
The analysis has shown an inherent link between quantum correlation and Nash equilibria. Interestingly, it appears that quantum correlation drives the system away from pure strategy Nash equilibria. In particular, it seems pure strategy Nash equilibria occur when there is classical correlation in the state for the chosen set of payoffs, either through the correlation between the players being quantum-classical, or entirely classical. This has been motivated by both analytical and numerical techniques, which give clear indications of how the differing correlation can affect the solutions for the Nash equilibria. By considering a variety of states which host differing types of quantum/classical correlation, the link between game theory and quantum physics has been highlighted.

Previously, there has been limited investigation on Nash equilibria in adversarial Bayesian games, as the standard technique used to find an equilibrium is by correlated equilibria, which admits a wider class of solutions. The primary benefit of that approach is that it is numerically easier to implement.  However the benefit is limited as it does not specifically find Nash equilibria solutions. It would be interesting to modify the algorithm implemented in this paper to examples where only correlated equilibria are considered, to understand how the set of strategy solutions differ, and whether similar results would be found. Ultimately, the analysis here relied on the game being constant-sum, enabling a minimax solution. However, it should be possible to modify the algorithm such that this restriction is lifted.

It is also an unknown issue whether the observed phenomena in terms of pure strategy Nash equilibria for quantum/classical states are exclusive to Bayesian games, or whether there is an inherent link between pure and mixed strategy Nash equilibria and quantum correlation in other game-theoretic frameworks. For example, understanding how Nash equilibria are affected in sequential games is an important future question. Additionally, whether discord motivates quantum advantage in these further games remains an open topic. 

\section{Conclusion}\label{s:Conclusion}
It has been found that there is a direct link between Nash equilibria and quantum correlation beyond entanglement, namely quantum discord. Specifically, there is a sufficient condition for constant-sum games, such that any Nash equilibria which can not be found using a pure strategy solutions, implies that there must be underlying quantum effects for both of the players. This implies that generally mixed strategy solutions are required in order to find valid Nash equilibria in constant-sum games when quantum correlations are present. 
This analysis will be useful for the design of future interacting quantum networks, in particular for determining the optimal strategies of the nodes, when individual parties perform local measurements in order to maximise the benefit in their respective situation.

\section{Acknowledgements}
The author thanks David Lowe for useful assistance developing the code for the {\it{Pseudo-Algorithm}}, and for critical comments with the development of this manuscript.

\section{Data Availability}
All figures can be produced with the information given in the text. The code for the {\it{Pseudo-Algorithm}} is available on reasonable request.

\section{Author Declaration}
There are no conflicts of interest.

\section*{References}
\bibliographystyle{ieeetr}
\bibliography{Literature_Review_References.bib}

\begin{thebibliography}{10}

\bibitem{aspect}
A.~{A}spect, P.~{G}rangier, and G.~{R}oger, ``{E}xperimental {R}ealization of
  {E}instein-{P}odolsky-{R}osen-{B}ohm {G}edankenexperiment: {A} {N}ew
  {V}iolation of {B}ell's {I}nequalities,'' {\em {P}hysical {R}eview
  {L}etters}, vol.~49, no.~2, pp.~91--94, 1982.

\bibitem{Passante:2011vz}
G.~Passante, O.~Moussa, D.~A. Trottier, and R.~Laflamme, ``Experimental
  detection of nonclassical correlations in mixed-state quantum computation,''
  {\em Physical Review A}, vol.~84, pp.~044302--, 10 2011.

\bibitem{Aslam:2023vu}
N.~Aslam, H.~Zhou, E.~K. Urbach, M.~J. Turner, R.~L. Walsworth, M.~D. Lukin,
  and H.~Park, ``Quantum sensors for biomedical applications,'' {\em Nature
  Reviews Physics}, vol.~5, no.~3, pp.~157--169, 2023.

\bibitem{Rarity:1990ul}
J.~G. Rarity and P.~R. Tapster, ``Experimental violation of bell's inequality
  based on phase and momentum,'' {\em Physical Review Letters}, vol.~64,
  pp.~2495--2498, 05 1990.

\bibitem{Pelucchi:2022ub}
E.~Pelucchi, G.~Fagas, I.~Aharonovich, D.~Englund, E.~Figueroa, Q.~Gong,
  H.~Hannes, J.~Liu, C.-Y. Lu, N.~Matsuda, J.-W. Pan, F.~Schreck, F.~Sciarrino,
  C.~Silberhorn, J.~Wang, and K.~D. J{\"o}ns, ``The potential and global
  outlook of integrated photonics for quantum technologies,'' {\em Nature
  Reviews Physics}, vol.~4, no.~3, pp.~194--208, 2022.

\bibitem{Hangleiter:2023te}
D.~Hangleiter and J.~Eisert, ``Computational advantage of quantum random
  sampling,'' {\em Reviews of Modern Physics}, vol.~95, pp.~035001--, 07 2023.

\bibitem{Madsen:2022wo}
L.~S. Madsen, F.~Laudenbach, M.~F. Askarani, F.~Rortais, T.~Vincent, J.~F.~F.
  Bulmer, F.~M. Miatto, L.~Neuhaus, L.~G. Helt, M.~J. Collins, A.~E. Lita,
  T.~Gerrits, S.~W. Nam, V.~D. Vaidya, M.~Menotti, I.~Dhand, Z.~Vernon,
  N.~Quesada, and J.~Lavoie, ``Quantum computational advantage with a
  programmable photonic processor,'' {\em Nature}, vol.~606, no.~7912,
  pp.~75--81, 2022.

\bibitem{quantumadvantage}
F.~{A}rute {\em et~al.}, ``{Q}uantum {S}upremacy using a {P}rogrammable
  {S}uperconducting {P}rocessor,'' {\em {N}ature}, vol.~574, pp.~505--510,
  2019.

\bibitem{Daley:2022vu}
A.~J. Daley, I.~Bloch, C.~Kokail, S.~Flannigan, N.~Pearson, M.~Troyer, and
  P.~Zoller, ``Practical quantum advantage in quantum simulation,'' {\em
  Nature}, vol.~607, no.~7920, pp.~667--676, 2022.

\bibitem{Azuma:2023tz}
K.~Azuma, S.~E. Economou, D.~Elkouss, P.~Hilaire, L.~Jiang, H.-K. Lo, and
  I.~Tzitrin, ``Quantum repeaters: From quantum networks to the quantum
  internet,'' {\em Reviews of Modern Physics}, vol.~95, pp.~045006--, 12 2023.

\bibitem{Brunner:2014vw}
N.~Brunner, D.~Cavalcanti, S.~Pironio, V.~Scarani, and S.~Wehner, ``Bell
  nonlocality,'' {\em Reviews of Modern Physics}, vol.~86, pp.~419--478, 04
  2014.

\bibitem{Tavakoli:2022uf}
A.~Tavakoli, A.~Pozas-Kerstjens, M.-X. Luo, and M.-O. Renou, ``Bell nonlocality
  in networks,'' {\em Reports on Progress in Physics}, vol.~85, no.~5,
  p.~056001, 2022.

\bibitem{Li:2023tc}
Z.~Li, K.~Xue, J.~Li, L.~Chen, R.~Li, Z.~Wang, N.~Yu, D.~S.~L. Wei, Q.~Sun, and
  J.~Lu, ``Entanglement-assisted quantum networks: Mechanics, enabling
  technologies, challenges, and research directions,'' {\em IEEE Communications
  Surveys \& Tutorials}, vol.~25, no.~4, pp.~2133--2189, 2023.

\bibitem{Pirandola:2014aa}
S.~Pirandola, ``Quantum discord as a resource for quantum cryptography,'' {\em
  Scientific Reports}, vol.~4, no.~1, p.~6956, 2014.

\bibitem{IP}
D.~{G}irolami, A.~M. {S}ouza, V.~{G}iovannetti, T.~{T}ufarelli, J.~G.
  {F}ilgueiras, R.~S. {S}arthour, D.~O. {S}oares {P}into, I.~S. {O}liveira, and
  G.~{A}desso, ``{Q}uantum {D}iscord {D}etermines the {I}nterferometric {P}ower
  of {Q}uantum {S}tates,'' {\em {P}hysical {R}eview {L}etters}, vol.~112,
  no.~21, p.~210401, 2014.

\bibitem{Lowe:2024aa}
A.~Lowe, ``Quantum advantage beyond entanglement in bayesian game theory,''
  {\em Journal of Physics A: Mathematical and Theoretical}, vol.~57, no.~6,
  p.~065303, 2024.

\bibitem{lowe}
A.~Lowe and I.~V. Yurkevich, ``The link between {F}isher information and
  geometric discord,'' {\em Low Temperature Physics}, vol.~48, p.~396, 2022.

\bibitem{Morimae:2014uh}
T.~Morimae, K.~Fujii, and J.~F. Fitzsimons, ``Hardness of classically
  simulating the one-clean-qubit model,'' {\em Physical Review Letters},
  vol.~112, pp.~130502--, 04 2014.

\bibitem{Knill:1998us}
E.~Knill and R.~Laflamme, ``Power of one bit of quantum information,'' {\em
  Physical Review Letters}, vol.~81, pp.~5672--5675, 12 1998.

\bibitem{Guha2021quantumadvantage}
T.~Guha, M.~Alimuddin, S.~Rout, A.~Mukherjee, S.~S. Bhattacharya, and M.~Banik,
  ``Quantum {A}dvantage for {S}hared {R}andomness {G}eneration,'' {\em
  {Quantum}}, vol.~5, pp.~569--586, 2021.

\bibitem{tadelis}
S.~{T}adelis, {\em {G}ame {T}heory: {A}n {I}ntroduction}.
\newblock Princeton University Press, 2012.

\bibitem{Ikeda:2023ab}
K.~Ikeda and A.~Lowe, ``Quantum protocol for decision making and verifying
  truthfulness among n-quantum parties: Solution and extension of the quantum
  coin flipping game,'' {\em IET Quantum Communication}, vol.~4, pp.~218--227,
  2024/05/29 2023.

\bibitem{nasheq}
J.~{N}ash, ``{N}on-{C}ooperative {G}ames,'' {\em {A}nnals of {M}athematics},
  vol.~54, no.~2, pp.~286--295, 1951.

\bibitem{gamebook}
J.~von {N}eumann, O.~{M}orgenstern, and A.~{R}ubinstein, {\em Theory of Games
  and Economic Behavior}.
\newblock Princeton University Press, 1944.

\bibitem{Farooqui:2016aa}
A.~D. Farooqui and M.~A. Niazi, ``Game theory models for communication between
  agents: a review,'' {\em Complex Adaptive Systems Modeling}, vol.~4, no.~1,
  p.~13, 2016.

\bibitem{Bayes_auctions}
G.~Christodoulou, A.~Kov\'{a}cs, and M.~Schapira, ``Bayesian combinatorial
  auctions,'' {\em J. ACM}, vol.~63, p.~11, 2016.

\bibitem{Pappa:2015aa}
A.~Pappa, N.~Kumar, T.~Lawson, M.~Santha, S.~Zhang, E.~Diamanti, and
  I.~Kerenidis, ``Nonlocality and conflicting interest games,'' {\em Physical
  Review Letters}, vol.~114, pp.~020401--, 01 2015.

\bibitem{Roy:2016aa}
A.~Roy, A.~Mukherjee, T.~Guha, S.~Ghosh, S.~S. Bhattacharya, and M.~Banik,
  ``Nonlocal correlations: Fair and unfair strategies in bayesian games,'' {\em
  Physical Review A}, vol.~94, pp.~032120--, 09 2016.

\bibitem{Brunner:2013aa}
N.~Brunner and N.~Linden, ``Connection between bell nonlocality and bayesian
  game theory,'' {\em Nature Communications}, vol.~4, no.~1, p.~2057, 2013.

\bibitem{Andronikos:2022aa}
T.~Andronikos, ``Conditions that enable a player to surely win in sequential
  quantum games,'' {\em Quantum Information Processing}, vol.~21, no.~7,
  p.~268, 2022.

\bibitem{Ikeda:2023aa}
K.~{I}keda, ``Quantum extensive-form games,'' {\em Quantum Information
  Processing}, vol.~22, no.~1, p.~66, 2023.

\bibitem{harsanyi}
J.~{H}arsanyi, ``Games with incomplete information played by ``bayesian''
  players, i--iii part i. the basic model.,'' {\em Management Science},
  vol.~14, no.~3, pp.~159--182, 1967.

\bibitem{PhysRevLett.82.1052}
D.~A. Meyer, ``{Quantum Strategies},'' {\em Phys. Rev. Lett.}, vol.~82,
  pp.~1052--1055, Feb 1999.

\bibitem{PhysRevA.65.022306}
A.~Iqbal and A.~H. Toor, ``Quantum mechanics gives stability to a {N}ash
  equilibrium,'' {\em Phys. Rev. A}, vol.~65, p.~022306, Jan 2002.

\bibitem{QgameRev}
F.~S. Khan, N.~Solmeyer, R.~Balu, and T.~S. Humble, ``Quantum games: a review
  of the history, current state, and interpretation,'' {\em Quantum Information
  Processing}, vol.~17, p.~309, 2018.

\bibitem{QgameRev1}
A.~P. Flitney and D.~Abbott, ``An introduction to quantum game theory,'' {\em
  Fluctuation and Noise Letters}, vol.~02, p.~175, 2002.

\bibitem{QGame_Locality}
C.~A. Melo-Luna, C.~E. Susa, A.~F. Ducuara, A.~Barreiro, and J.~H. Reina,
  ``Quantum locality in game strategy,'' {\em Scientific Reports}, vol.~7,
  p.~44730, 2017.

\bibitem{PHOENIX2020126299}
S.~Phoenix, F.~Khan, and B.~Teklu, ``Preferences in quantum games,'' {\em
  Physics Letters A}, vol.~384, no.~15, p.~126299, 2020.

\bibitem{cite-key}
M.~Banik, S.~S. Bhattacharya, N.~Ganguly, T.~Guha, A.~Mukherjee, A.~Rai, and
  A.~Roy, ``Two-{Q}ubit {P}ure {E}ntanglement as {O}ptimal {S}ocial {W}elfare
  {R}esource in {B}ayesian {G}ame,'' {\em {Quantum}}, vol.~3, 2019.

\bibitem{10.1063/5.0204288}
A.~Banerjee, P.~Bej, A.~Mukherjee, S.~G. Naik, M.~Alimuddin, and M.~Banik,
  ``{When Mei-Gu Guan's 1960 postmen get empowered with Bell's 1964 nonlocal
  correlations: Nonlocal advantage in vehicle routing problem},'' {\em APL
  Quantum}, vol.~1, p.~036105, 07 2024.

\bibitem{discord}
H.~{O}llivier and W.~H. {Z}urek, ``{Q}uantum {D}iscord: {A} {M}easure of the
  {Q}uantumness of {C}orrelations,'' {\em {P}hysical {R}eview {L}etters},
  vol.~88, no.~1, p.~017901, 2001.

\bibitem{discord1}
L.~{H}enderson and V.~{V}edral, ``{C}lassical, quantum and total
  correlations,'' {\em {J}ournal of {P}hysics {A}: {M}athematical and
  {G}eneral}, vol.~34, no.~35, pp.~6899--6905, 2001.

\bibitem{Werlang:2009aa}
T.~Werlang, S.~Souza, F.~F. Fanchini, and C.~J. Villas~Boas, ``Robustness of
  quantum discord to sudden death,'' {\em Physical Review A}, vol.~80,
  pp.~024103--, 08 2009.

\bibitem{PhysRevResearch.6.023248}
A.~Ulibarrena, A.~Sopena, R.~Brooks, D.~Centeno, J.~Ho, G.~Sierra, and
  A.~Fedrizzi, ``Photonic implementation of the quantum morra game,'' {\em
  Phys. Rev. Res.}, vol.~6, p.~023248, Jun 2024.

\bibitem{doi:10.1126/science.abe8770}
H.-S. Zhong, H.~Wang, Y.-H. Deng, M.-C. Chen, L.-C. Peng, Y.-H. Luo, J.~Qin,
  D.~Wu, X.~Ding, Y.~Hu, P.~Hu, X.-Y. Yang, W.-J. Zhang, H.~Li, Y.~Li,
  X.~Jiang, L.~Gan, G.~Yang, L.~You, Z.~Wang, L.~Li, N.-L. Liu, C.-Y. Lu, and
  J.-W. Pan, ``Quantum computational advantage using photons,'' {\em Science},
  vol.~370, no.~6523, pp.~1460--1463, 2020.

\bibitem{chsh}
J.~F. Clauser, M.~A. Horne, A.~Shimony, and R.~A. Holt, ``Proposed experiment
  to test local hidden-variable theories,'' {\em Physical Review Letters},
  vol.~23, pp.~880--884, 10 1969.

\bibitem{Cleve:2004tw}
R.~Cleve, P.~Hoyer, B.~Toner, and J.~Watrous, ``Consequences and limits of
  nonlocal strategies,'' in {\em Proceedings. 19th IEEE Annual Conference on
  Computational Complexity, 2004.}, pp.~236--249, 2004.

\end{thebibliography}

\end{document}